\documentclass{article}[12pt]
\usepackage{epsfig,dcolumn,bm}

\newcommand{\dmu}{\partial_\mu}

\newcommand{\lc}{{\cal L}}

\newcommand{\sqd}{\sqrt{2}}

\newcommand{\be}{\begin{eqnarray}}
\newcommand{\ee}{\end{eqnarray}}
\newcommand{\nn}{\nonumber}

\title{Testing the predicted dynamically generated hidden charm $X$ scalar state through the $D\bar D$ invariant mass spectrum and the radiative decay of the $\psi(3770)$}

\author{D. Gamermann$^1$, E. Oset$^1$, B. S. Zou$^2$ \\ 
{\small{\it $^1$Hadronic and Nuclear Theory Group-IFIC, CSIC-Universidad de Valencia - Spain}}\\
{\small{\it $^2$IHEP and Theoretical Physics Center for Science Facilities - CAS - China}}
}

\begin{document}

\maketitle

\abstract{In this work we present our model that dynamically generates resonances from charmed and light pseudoscalar and vector mesons. The model generates a rich spectrum of charmed and hidden charm scalar and axial resonances, most of them which can be associated with observed states. Other states are predicted, in particular a hidden charm scalar resonance with mass close to the $D\bar D$ threshold. For this resonance we present calculations which show that an observed enhancement in the $D\bar D$ mass spectrum measured by Belle in reactions producing $D\bar D J/\psi$ in the final state could be originated by this predicted state. We also present theoretical calculations on the radiative decay width of the $\psi(3770)$ into this state. The predicted branching fractions suggest that this predicted state could be observed by BES with statistics of one year run in the $\psi(3770)$ peak.}

\section{Introduction}

We have developed a phenomenological model in order to construct the interaction of pseudoscalars and vector mesons including charmed mesons \cite{meusca,meuax}. The model extends commonly used $SU(3)$ Lagrangians to $SU(4)$, including in this way the charmed mesons and the $J/\psi$. In this model $SU(4)$ symmetry is explicitly broken by the assumption that the underlying interaction in between the hadronic currents is driven by vector mesons, in this way whenever a charmed vector meson should be exchanged, the term in the Lagrangian corresponding to this interaction is suppressed by a factor which takes into account the heavy vector meson mass. This model reproduces most of the results obtained combining heavy quark symmetry with chiral symmetry \cite{lutz1,lutz2,guo1,guo2}, but it also contains results for the hidden charm sector and for the interaction of light vector mesons with heavy pseudoscalars which could not be grasped from the heavy quark symmetry approach.

Apart from generating many resonances that can be associated with observed states from numerous experiments (BaBar, Belle, CLEO, CDF, ...), the model also makes predictions of other states that have not been observed. In particular, it predicts a narrow hidden charm scalar state with mass close to the $D\bar D$ threshold. In this presentation we calculate the $D\bar D$ mass spectrum from the decay of the dynamically generated hidden charm scalar state and compare it to the results from the $D\bar D$ mass spectrum measured by Belle from the reaction $e^+e^-\rightarrow J/\psi D\bar D$ \cite{belle1}. Our results fairly describe the data, giving some evidence for its existence. We also performed a calculation for the radiative decay width of the $\psi(3770)$ into this state and the results suggest that this radiative decay could be observed by the BES experiment with one year statistics in the $\psi(3770)$ peak.

The work is organized as follows, in the next section we present our model and the spectrum it generates. In section 3 we show the calculation of the $D\bar D$ invariant mass spectrum and in section 4 we show the calculation for the radiative decay width of the $\psi(3770)$. Last section contains an outlook and our final remarks.


\section{Phenomenological Model}

To construct our Lagrangians we first write down fields for the 15-plets of $SU(4)$, plus a singlet, that represent the pseudoscalar and vector meson fields:

\be
\Phi=\left(
\begin{array}{cccc}
 \frac{\eta }{\sqrt{3}}+\frac{\pi^0}{\sqrt{2}}+\frac{\eta'
   }{\sqrt{6}} & \pi ^+ & K^+ & \overline{D}^0 \\& & & \\
 \pi ^- & \frac{\eta }{\sqrt{3}}-\frac{\pi
   ^0}{\sqrt{2}}+\frac{\eta'}{\sqrt{6}} & K^0 & D^- \\& & & \\
 K^- & \overline{K}^0 & \sqrt{\frac{2}{3}} \eta'-\frac{\eta
   }{\sqrt{3}} &  {D_s}^- \\& & & \\
 D^0 & D^+ &  {D_s}^+ & \eta _c
\end{array}
\right)  \nn \\
\cal{V}_\mu=\left( \begin{array}{cccc}
{\rho_\mu^0 \over \sqd}+{\omega_\mu \over \sqd} & \rho^+_\mu & K^{*+}_\mu & \bar D^{*0}_\mu \\ & & & \\
\rho^{*-}_\mu & {-\rho^0_\mu \over \sqd}+{\omega_\mu \over \sqd} & K^{*0}_\mu & D^{*-}_\mu \\& & & \\
K^{*-}_\mu & \bar K^{*0}_\mu & \phi_\mu & D_{s\mu}^{*-} \\& & & \\
D^{*0}_\mu & D^{*+}_\mu & D_{s\mu}^{*+} & J/\psi_\mu \\ \end{array} \right). \nn
\ee
Note that these fields differ from those used in \cite{meusca,meuax} because of the inclusion of $\eta$-$\eta'$
and $\omega$-$\phi$ mixing.

For each one of these fields a current is defined:

\be
J_\mu&=&(\dmu \Phi)\Phi-\Phi\dmu\Phi \\
\cal{J}_\mu&=&(\dmu \cal{V}_\nu)\cal{V}^\nu-\cal{V}_\nu\dmu \cal{V}^\nu. 
\ee

The Lagrangians are constructed by coupling these currents:

\be
\lc_{PPPP}&=&{1\over12f^2}Tr({J}_\mu {J}^\mu+\Phi^4 M) \label{lagsca}\\
\lc_{PPVV}&=&-{1\over 4f^2}Tr\left(J_\mu\cal{J}^\mu\right). \label{lagax}
\ee
The matrix $M$ in (\ref{lagsca}) is diagonal and given by: $M=diagonal(m_\pi^2,m_\pi^2,2m_K^2-m_\pi^2, 2m_D^2-m_\pi^2)$.

These Lagrangians are $SU(4)$ symmetric but, since $SU(4)$ symmetry is badly
broken in nature, we will break the $SU(4)$ symmetry of the Lagrangians in the following way:
Assuming vector-meson dominance we recognize that the interaction behind our Lagrangians is the exchange of a vector
meson in between the two hadronic currents. If the initial and final pseudoscalars (and vector-mesons), in a given process,
have different charm quantum number, it means that the vector-meson exchanged in such process is a charmed meson,
and hence a heavy one. In these cases we suppress the term by the factor $\gamma=m_L^2/m_H^2$ where $m_L$ is the typical value of a light vector-meson mass (800 MeV) and $m_H$ the typical value of the heavy vector-meson mass (2050 MeV). We also 
suppress, in the interaction of $D$-mesons the amount of the interaction which is driven by a $J/\psi$ exchange by the factor
$\psi=m_L^2/m_{J/\psi}^2$. Another source of symmetry breaking will be the meson decay constant $f$ appearing in the Lagrangian.
For light mesons we use $f=f_\pi=93$ MeV but for heavy ones $f=f_D=165$ MeV.

From each Lagrangian one gets three level amplitudes for any two meson initial and final state that expand a coupled channel space. These amplitudes are projected in s-wave and collected in a matrix $V$ that we plug as kernel to solve the scattering equation:

\be
T&=&V+VGT. \label{bseq}
\ee
In this equation $G$ is a diagonal matrix with each one of its elements given by the loop function for each channel in the coupled channel space. For channel $i$ with mesons of masses $m_1$ and $m_2$ $G_{ii}$ is given by:

\be
G_{ii}&=&{1 \over 16\pi ^2}\biggr( \alpha _i+Log{m_1^2 \over \mu ^2}+{m_2^2-m_1^2+s\over 2s}
  Log{m_2^2 \over m_1^2} 
 + {p\over \sqrt{s}}\Big( Log{s-m_2^2+m_1^2+2p\sqrt{s} \over -s+m_2^2-m_1^2+
  2p\sqrt{s}}\nn\\
&+&Log{s+m_2^2-m_1^2+2p\sqrt{s} \over -s-m_2^2+m_1^2+  2p\sqrt{s}}\Big)\biggr)
\ee
where $p$ is the three momentum of the two mesons in the center of mass frame. The two parameters $\mu$ and $\alpha$ are not independent, we fix $\mu$=1500 MeV and change $\alpha$ to fit our results within reasonable values in the natural range \cite{hyodo}. We actually use two different alphas as free parameters, one for loops with only light mesons in the channel $i$ and another one for loops with heavy particles that we call $\alpha_H$.

The imaginary part of the loop function ensures that the T-matrix is unitary, and since this imaginary part is known, it is possible to do an analytic continuation for going from the first Riemann sheet to the second one. Possible physical states (resonances) are identified as poles in the T-matrix calculated in the second Riemann sheet for the channels which have the threshold below the resonance mass.

At first we studied the $SU(3)$ structure of the interaction. A 15-plet of $SU(4)$ breaks down into four multiplets of the lower $SU(3)$ symmetry: an octet and a singlet with null charm quantum number and an antitriplet and a triplet with positive and negative charm, respectively. Apart from the octets and singlets with null charm coming from the scattering of two light octets, one can also expect hidden charm singlets coming from the scattering of the heavy triplets with the antitriplets. This results in one hidden charm scalar resonance and two axial ones. In the open charm sector one can expect to generate poles corresponding to antitriplets and sextets and their respective anti-multiplets.

For the scalars, in the open charm sector, an antitriplet is generated \cite{meusca}. It has two isospin members that can be identified with the $D_{s0}^*(2317)$ and $D_0^*(2400)$. The sextet generated comes out too broad, so one cannot expect to be able to identify it experimentally.

Still in the scalar sector one can observe a hidden charm resonance coming mainly from the interaction between the $D$ and $\bar D$ mesons. This resonance appears around 3.7 GeV and is very narrow if one does not consider the $\eta-\eta\prime$ mixing. Considering this mixing, this resonance acquires a width around 40 MeV, still rather narrow if compared with other hadronic resonances.

The spectrum for the axial resonances is richer, since in this case one can distinguish between a light pseudoscalar scattering with a heavy vector and a light vector scattering with a heavy pseudoscalar. So, in the open charm sector two antitriplets are generated, one of them was identified with the $D_{s1}(2460)$ and $D_1(2430)$ while the other one with the $D_{s1}(2536)$ and $D_1(2420)$ resonances \cite{meuax}. Furthermore in the present case, one of the two sextets generated in this sector is very broad but the second one has relatively smaller width ($\sim$200 MeV), so in principle these exotic states could also be observed.

For the hidden charm sector two singlets can be identified, they are states of defined G-parity. The positive state is identified with the $X(3872)$ state, while the negative G-parity state is a prediction of our model. In \cite{meuax} the subtraction constant in the loop function was set to -1.55, in this case the two resonances appear around 3.84 GeV and are degenerate up to 3 MeV. If one takes a bigger subtraction constant, around -1.3, the two states are nearly degenerated ($M_+-M_-<0.5$ MeV) around the experimental value of 3.87 GeV. In this situation only experiments that can select definite G(or C)-parity would be able to differentiate between the two.

\section{The $D\bar D$ invariant mass spectrum}

\begin{figure}[h]
\begin{center}
\includegraphics[width=6cm]{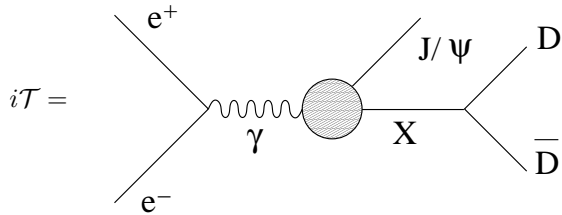} 
\put(-210,40){$i{\cal T}=$}
\end{center}
\caption{Feynman diagram for the process $e^+e^-\rightarrow J/\psi D\bar D$} \label{reac}
\end{figure}

Let us assume that in the reaction $e^+e^-\rightarrow J/\psi D\bar D$ the $D\bar D$ pair comes from a resonance, as shown in figure \ref{reac}.

If the $D\bar D$ invariant mass is close to threshold, the only part of the amplitude $\cal T$ which is strongly energy dependent is the $X$ propagator. In this case $\cal T$ would be proportional to the $X$ propagator and depend only on the $D\bar D$ invariant mass. The differential cross section would then be given by \cite{meucross}:

\begin{eqnarray}
{d\sigma\over dM_{inv}(D\bar D)}&=&{1\over(2\pi)^3}{m_e^2\over s\sqrt{s}} |\overrightarrow k| |\overrightarrow p| |{\cal T}|^2 \label{dcross}
\end{eqnarray}
where $s$ is the center of mass energy of the electron positron pair squared, $|\overrightarrow p|$ is the $J/\psi$ three momentum in the center of mass frame of the $J/\psi$ with the $D\bar D$ system and $|\overrightarrow k|$ is the $D$ meson three momentum in the center of mass frame of the $D\bar D$ system.

If the $X$ is a genuine resonance its propagator is of the Breit-Wigner type. In our approach $X$ is a dynamically generated resonance, so the information of its propagator is contained in the scattering $T$-matrix calculated from eq. (\ref{bseq}) and we use for $\cal T$ of eq. (\ref{dcross}) the $D\bar D$ channel of the $T$-matrix.

We performed a standard $\chi^2$ test in order to compare our theoretical calculation \cite{meucross} with experimental data from \cite{belle1}. The results are shown in table \ref{tabres} for different values of the parameter $\alpha_H$ varied within its natural range.

\begin{table}[h]
\begin{center}
\caption{Results of $M_X$ and $\chi^2$ for different values of $\alpha_H$.} \label{tabres}
\begin{tabular}{c|cc}
\hline
$\alpha_H$ & $M_X$(MeV) & $\chi^2 \over d.o.f$ \\
\hline
\hline
-1.4   & 3702 & 0.96 \\
-1.3   & 3719 & 0.85 \\
-1.2  & 3728 & 0.92 \\
-1.1   & Cusp & 1.11 \\
\hline
\end{tabular}
\end{center}
\end{table}

The results show good agreement of our approach to the data \cite{meucross} giving some support to the existence of this scalar $X$ resonance.

\section{The radiative decay of the $\psi(3770)$ into the scalar $X$ resonance}

Here we study the following radiative decay: 

\be
\psi(P,\epsilon(P))\rightarrow X(Q)+\gamma(K,\epsilon(K)) \label{decay}
\ee
where $\psi$ is the $\psi(3770)$ and $X$ is the hidden charm scalar dynamically generated resonance $X(3700)$.

\begin{figure}[h]
\begin{center}
\begin{tabular}{cc}
\includegraphics[width=7.5cm,angle=-0]{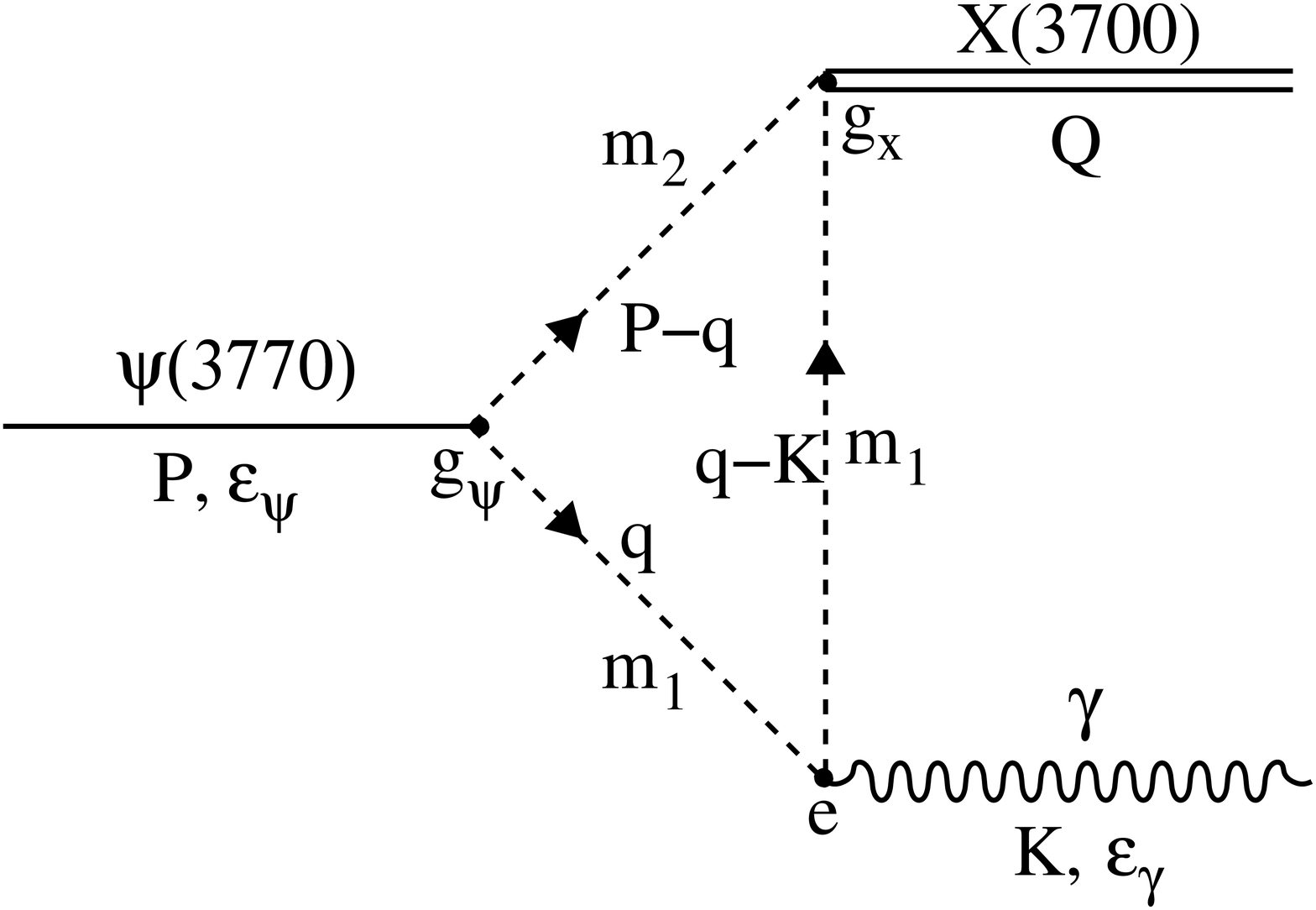} &\includegraphics[width=7.5cm,angle=-0]{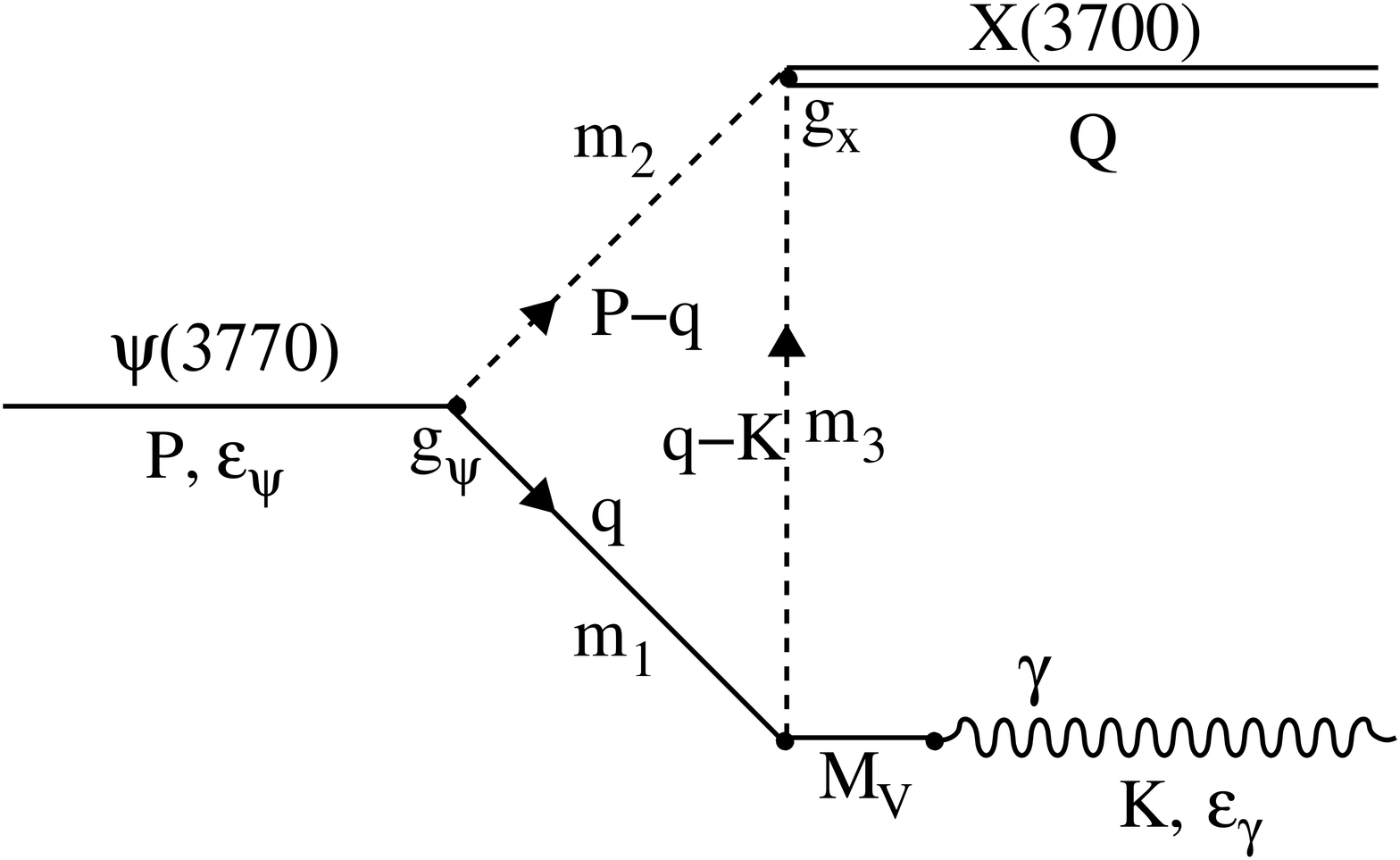} 
\end{tabular}
\caption{Diagrams needed for the evaluation of the amplitude for the radiative decay.} \label{diagsdecay}
\end{center}
\end{figure}

Two diagrams should be considered in order to evaluate the full gauge invariant amplitude for this precess. These diagrams are represented in figure \ref{diagsdecay}.

As explained in \cite{meudecay} other diagrams needed for the evaluation of the full gauge invariant amplitude are related to the ones shown in figure \ref{diagsdecay} and do not need to be explicitly calculated. The expressions for the diagrams can also be found in \cite{meudecay} with their derivation explained in detail.

In table \ref{tabdecay} we show results for the radiative decay width of the $\psi(3770)$ into the $X$ resonance for different values of the parameter $\alpha_H$.

\begin{table}[h]
\begin{center}
\caption{Results for the radiative decay width of the $\psi(3770)$ calculated for different values of $\alpha_H$.} \label{tabdecay}
\begin{tabular}{c|c}
\hline
$\alpha_H$& $\Gamma$ [KeV]\\
\hline
\hline
-1.40&1.97 \\
\hline
-1.35&1.09 \\
\hline
-1.30&0.79 \\
\hline
-1.25&0.55 \\
\hline
\end{tabular}
\end{center}
\end{table}

The results correspond to branching fractions of the same order of magnitude as other observed radiative decays like $\phi\rightarrow a_0(980)\gamma$ or $\phi\rightarrow f_0(980)\gamma$ which are observed reactions and proceed through similar loops but involving kaons instead of $D$ mesons. 

The BEPC-II facility is expected to produce $3.8\times 10^7$ $\psi(3770)$ events in one year of run, this would correspond to about 1000 events  into $X(3700)\gamma$ for $\Gamma_{\psi\rightarrow X\gamma}\sim 1$ KeV, which should be enough statistic in order to observe a clear peak, disregarding any technical problems that are beyond our reach.

\section{Outlook}

In this presentation we have reviewed the main aspects of the dynamical generation of resonances with charm and hidden charm. Our model starts by considering a $SU(4)$ symmetrical Lagrangian. The $SU(4)$ symmetry is then broken down to $SU(3)$ with the assumption of vector meson dominance. From the resulting Lagrangian, three level amplitudes are obtained for the interaction of pairs of mesons. With this amplitudes a multi-channel $T$-matrix is calculated in the whole complex plane. Poles in this $T$-matrix are identified with resonances, most of them known experimentally, but some are predictions which should be tested against experiment.

A particular prediction of the model is a hidden charm scalar state with mass close to the $D\bar D$ threshold. We have focused in this state an used it as input to calculate the invariant mass spectrum of a $D\bar D$ pair generated in the reaction $e^+e^-\rightarrow J/\psi D\bar D$ which has been measured by Belle. The theoretical calculations fairly describe the data, giving us further encouragement to investigate the consequences of the existence of this state. A calculation of the radiative decay width of the $\psi(3770)$ into this state suggests that within a few years there may be enough statistics available from the BES experiment in order to identify a clear peak for this decay and thus confirm the existence of this state.

We hope to encourage further experimental studies in order to confirm (or discard) the existence of this state.

\section*{Acknowledgments}

This work is 
partly supported by DGICYT contract number
FIS2006-03438. We acknowledge the support of the European Community-Research 
Infrastructure Integrating Activity
"Study of Strongly Interacting Matter" (acronym HadronPhysics2, Grant Agreement
n. 227431) under the Seventh Framework Programme of EU. 
Work supported in part by DFG (SFB/TR 16, "Subnuclear Structure of Matter").


\end{document}